# The Evolution of the *GAL*actose Utilization Pathway in Budding Yeasts

Marie-Claire Harrison[1], Abigail L. LaBella[1], Chris Todd Hittinger[2,*], & Antonis Rokas[1,*]

[1] *Department of Biological Sciences, Vanderbilt University, TN, USA*

[2] *Laboratory of Genetics, DOE Great Lakes Bioenergy Research Center, Wisconsin Energy Institute, Center for Genomic Science Innovation, J.F. Crow Institute for the Study of Evolution, University of Wisconsin-Madison, Madison, WI, USA*

*Authors for correspondence: cthittinger@wisc.edu and antonis.rokas@vanderbilt.edu

**Abstract**

The Leloir galactose utilization or *GAL* pathway of budding yeasts, including that of the baker's yeast *Saccharomyces cerevisiae* and the opportunistic human pathogen *Candida albicans*, breaks down the sugar galactose for energy and biomass production. The *GAL* pathway has long served as a model system for understanding how eukaryotic metabolic pathways, including their modes of regulation, evolve. More recently, the physical linkage of the structural genes *GAL1*, *GAL7*, and *GAL10* in diverse budding yeast genomes has been used as a model for understanding the evolution of gene clustering. In this opinion, we summarize exciting recent work on three different aspects of this iconic pathway's evolution: gene cluster organization, *GAL* gene regulation, and the population genetics of the *GAL* pathway.

**Highlights**

- The *GAL* pathway of budding yeasts is a powerful model for inferring the evolutionary principles guiding the evolution of eukaryotic metabolic and genetic pathways.

- The *GAL* pathway exhibits substantial variation in its genomic organization across budding yeast species, and a few different mechanisms have driven the evolution of this organization.

- At least two distinct modes of regulation of the *GAL* pathway are known in the budding yeast subphylum; there is likely substantial variation between species that use these modes of regulation, and other yet-to-be-discovered modes of regulation likely exist in the subphylum.

- Population genomic studies have revealed extensive genetic variation, including alternative and highly distinct *GAL* gene network variants, within budding yeast species, suggesting that yeast populations are subject to varied selection for the utilization of the galactose present in different environments and conditions.

**Introduction**

Galactose is a monosaccharide that is abundant in nature and is found in many forms. For example, galactose is a component of the disaccharide lactose, a main ingredient of dairy products, as well as the trisaccharide raffinose and disaccharide melibiose, which are both common in grains and other plants (Acosta & Gross, 1995). Galactose is also often found in the form of glycolipids and glycoproteins, which are produced by cells and are critical for a wide variety of cellular functions (Coelho et al. 2015, Acosta & Gross, 1995) (Box 1). The assimilation of galactose has been extensively characterized both in humans (Coelho et al. 2015) and in the baker's yeast *Saccharomyces cerevisiae* (Sellik et al. 2008). The enzymatic products of three genes, *GAL1*, *GAL7*, and *GAL10* in budding yeasts, or *GALK*, *GALT*, *GALM*, and *GALE* in humans, are responsible for converting galactose into glucose-1-phosphate (Figure 1; Sellik et al. 2008). Glucose-1-phosphate is then isomerized to glucose-6-phosphate by Pgm1p or Pgm2p, which can then be used to generate energy for the cell via glycolysis.

Through multiple now-classic studies, the Leloir galactose utilization or *GAL* pathway of budding yeasts has been established as a model system for understanding the function and evolution of eukaryotic metabolic pathways and their regulation (Johnston 1987, Hittinger et al. 2004, Hittinger et al. 2010, Slot & Rokas, 2010, Dalal et al. 2016). We will start by introducing the classic work on the evolution of the *GAL* pathway from comparative studies of *S. cerevisiae* and *Candida albicans*. Then we will synthesize several discoveries in the last few years that have significantly enriched our understanding of how this iconic pathway evolved. These results include the discovery of repeated instances of wholesale pathway loss and of reacquisition via horizontal gene transfer (HGT), extensive variation in sequence and function within species, and diversity in levels and modes of pathway regulation and genomic organization.

---

**Box 1. Glycosylation and the *GAL* pathway**

Glycosylation, the enzymatic process by which glycans are attached to protein molecules, is an essential part of creating mature proteins in eukaryotic cells and enables many cellular functions. Galactose and an amino sugar derivative of galactose, *N*-acetylgalactosamine (GalNac), are parts of many glycans. Genetic mutations that affect glycosylation are associated with many human disorders, including inflammatory and autoimmune diseases, impaired kidney function, and decreased immune recognition of microbes (Reily et al. 2019). The importance of glycosylation is also highlighted by sugar starvation, in which cells create proteins with immature glycoproteins that lead to aggregation and endoplasmic reticulum stress (Sasaoka et al. 2018). There are two main types of glycosylation: *N*-linked and *O*-linked. *O*-linked glycosylation involves a serine or threonine residue attached to a GalNac, whereas *N*-linked glycosylation involves an asparagine residue attached to *N*-acetylglucosamine (GlcNac), an amino acid derivative of glucose. Other sugars such as glucose, mannose, and galactose, are often added to these to create larger glycans. Fascinatingly, studies in mammals suggest that the *GAL* pathway itself may also be critical for glycosylation: trace amounts of galactose in human embryonic kidney cells have been shown to rescue sugar-starved cells from endoplasmic reticulum stress and apoptosis while making mature glycoprotein production much more effective than adding glucose and mannose (Sasaoka et al. 2018). This phenomenon is likely due to the increased flux into fructose-6-phosphate, an intermediate that can be then used for *N*-linked glycosylation, through the *GAL* pathway in these cells (Sasaoka et al. 2018).

---

**The classical view: evolution of the *GAL* pathway in *S. cerevisiae* and *C. albicans***

Comparison of the *GAL* genes, the *GAL* pathway's genomic organization, and regulation between *C. albicans* and *S. cerevisiae* shows that there is substantial variation between the two species. While the structural genes are functional orthologs of each other and have the same order and orientation in the two species, the *C. albicans* cluster also contains the genes *GAL102* and *ORF-X* (Figure 1). *GAL102* encodes a glucose-4,6-dehydratase and *ORF-X* encodes a transporter. Although transporters of galactose (encoded by *GAL2* and other *HXT* genes for *S. cerevisiae* and by *HXT* genes in *C. albicans*; (Van Ende et al. 2019; Brown et al. 2009)) and the *PGM1/PGM2* structural genes are important parts of both the *S. cerevisiae* and *C. albicans GAL* pathways, none of these genes are parts of their gene clusters.

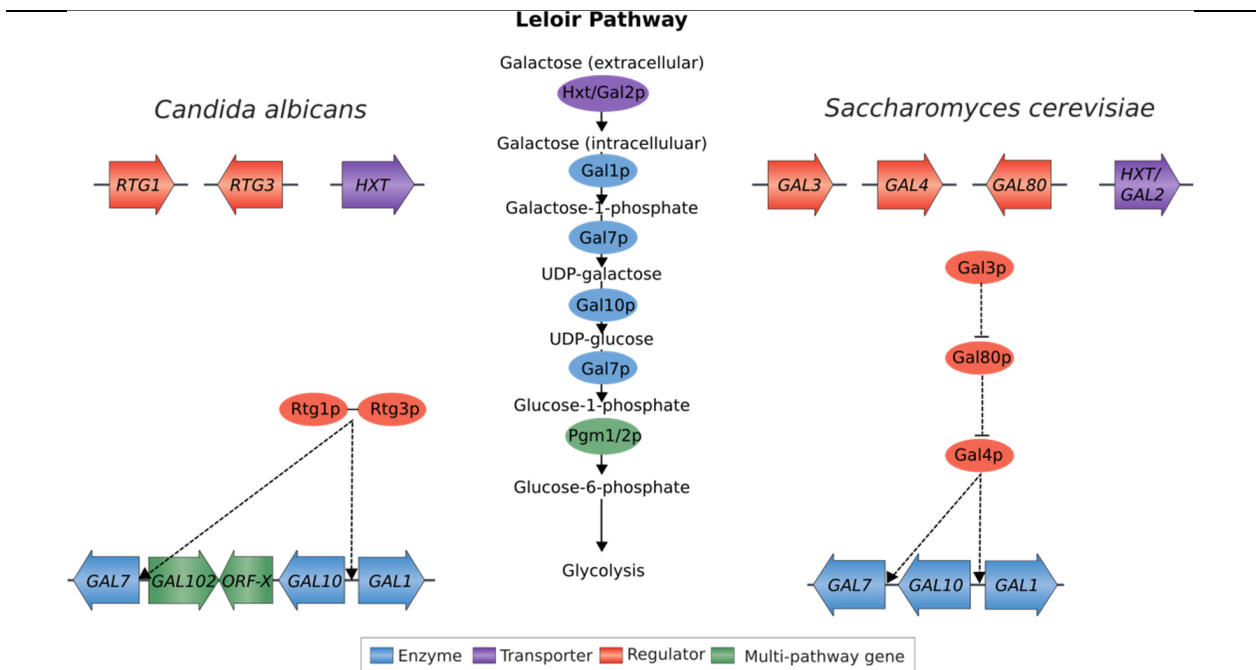

**Figure 1.** Comparison of genomic organization, function, and mode of regulation of the *GAL* pathway in the model organisms *C. albicans* (left panel) and *S. cerevisiae* (right

panel). Although *GAL102* and *ORF-X* are nested within the *GAL* gene cluster in *C. albicans*, their functions are not known to be related to galactose assimilation. Information displayed in the figure based on: Sellik et al. 2008, Dalal et al. 2016, and Haase et al. 2021.

---

The regulation of the *GAL* pathway also varies between the two organisms. In *S. cerevisiae*, the pathway is regulated by the Zn-binuclear cluster transcription factor Gal4p, which binds to a highly enriched regulatory motif whose consensus sequence is 5'-CGG-$N_{11}$-CCG-3'. Gal4p is repressed by Gal80p in the absence of galactose. In the presence of galactose, the repression of Gal80p is removed by Gal3p, and Gal4p induces transcription of the *GAL1*, *GAL7*, and *GAL10* genes. This leads to an "on-and-off switch" or bimodal mode of regulation, with strong suppression of the pathway when galactose is absent and a ~900-fold induction when galactose is detected (Dalal et al. 2016). In contrast, the *GAL* pathway of *C. albicans* is regulated by the heterodimeric helix-loop-helix transcription factors Rtg1p and Rtg3p. These transcription factors bind to a different binding motif, whose consensus sequence is 5'-TGYAACGTTRCA-3'. The basal rate of transcription of *GAL* genes in *C. albicans* is higher, and the induction is more graded, with a ~12-fold induction in the presence of galactose (Dalal et al. 2016). The differences between the two species in *GAL* pathway genomic organization and regulation are likely adaptations to the species' different ecologies and environments.

**Budding yeasts are a model lineage for studying the evolution of metabolic pathways**

There are ~1,200 known species of budding yeasts that belong to the subphylum Saccharomycotina, one of the three subphyla in the phylum Ascomycota (Shen et al. 2018, Shen et al. 2020, Li et al. 2021, Kurtzman et al. 2011). Advances in genome sequencing technologies have led to the sequencing of the genomes of hundreds of species across the subphylum, allowing for a greater understanding of how these organisms, their genes, and their metabolic traits evolved. The two biggest efforts to date have been the genome sequencing of 16 diverse species of biotechnologically important yeast species by a Joint Genome Institute-led effort (Riley et al. 2016) and the genome sequencing of 220 species (including 24 from the RIKEN Institute in Japan) led by the Y1000+ Project (http://y1000plus.org/), which placed emphasis on sequencing at least one representative species from each genus of budding yeasts (Shen et al. 2018). These efforts have led to a robust higher-level phylogeny of budding yeasts and the identification of 12 major lineages or clades (Shen et al. 2018); *S. cerevisiae* belongs to the family Saccharomycetaceae, while *C. albicans* belongs to the CUG-Ser1 clade (so called because the CUG codon encodes for serine rather than leucine in this clade (Krassowski et al. 2018)).

This richness of genomic and phylogenetic data is complemented by extensive aggregated metabolic and ecological trait data for a broad and representative set of budding yeast species. These include qualitative growth data on 44 substrates and environmental isolation data for up to 50 environments for 784 species, and quantitative growth rate data for galactose, mannose, and glucose for 258 species (Kurtzman et al. 2011, Opulente et al. 2018, Labella et al. 2021). These genomes, metabolic and ecological data, and species phylogeny, coupled with the availability of strains from all known species, have provided an unprecedented resource that has allowed for deeper analysis of metabolic pathways, including the *GAL* pathway.

**Evolution of genomic organization of the *GAL* pathway across the budding yeast subphylum**

Approximately half of the ~350 species with available genomes studied to date species contain the *GAL1, GAL7,* and *GAL10* genes and can grow on galactose (Shen et al., 2018, Opulente et al. 2018). Furthermore, out of the 174 species that grow on galactose, 127 species have the *GAL1, GAL7,* and *GAL10* genes clustered, and 23 additional species that do not grow on galactose also have the genes clustered, indicating they still may use the pathway or that it was recently inactivated (Shen et al. 2018, Labella et al. 2021). The 47 species that grow on galactose without having the genes clustered are mostly in clades more divergent from the CUG-Ser1 and Saccharomycetaceae clades, such as the Dipodascaceae/Trichomonascaceae clade, which contains the genera *Blastobotrys* and *Yarrowia*. However, a few species in Saccharomycetaceae that are descendants of an ancient whole genome duplication event (Wolfe & Shields 1997; Marcet-Houben & Gabaldón 2015), such as *Vanderwaltozyma polyspora* (Slot & Rokas, 2010), also have a functional *GAL* pathway but lack a cluster; this genome duplication event was followed by extensive loss of duplicate genes, such that some *GAL* genes are now found in one ohnologous genomic region and the rest are found in the other.

The clustering of *GAL1, GAL7,* and *GAL10* has evolved multiple times in fungi. For example, *Cryptococcus* basidiomycetous yeasts also have a *GAL* cluster, but phylogenetic analysis suggests that this cluster evolved independently, and its organization differs from the budding yeast *GAL* clusters (Slot & Rokas, 2010). In budding yeasts, clustering of the *GAL* genes originated at least twice: once in the common ancestor of *S. cerevisiae* and *C. albicans* and another time in *Lipomyces* and relatives (Haase et al. 2021, Figure 2). *Lipomyces* species have

*GAL10* next to *GAL7*, instead of *GAL1*, and they often have two copies of *GAL1* with an uncharacterized transcription factor between them that is homologous to *ARA1*, the L-arabinose regulatory transcription factor in *Trichoderma reesei* (Haase et al. 2021, Benocci et al. 2017). The repeated origin of the clustering of *GAL1*, *GAL7*, and *GAL10* in budding yeasts and other fungi supports the hypothesis that this genomic organization may be selectively advantageous in certain conditions (Box 2).

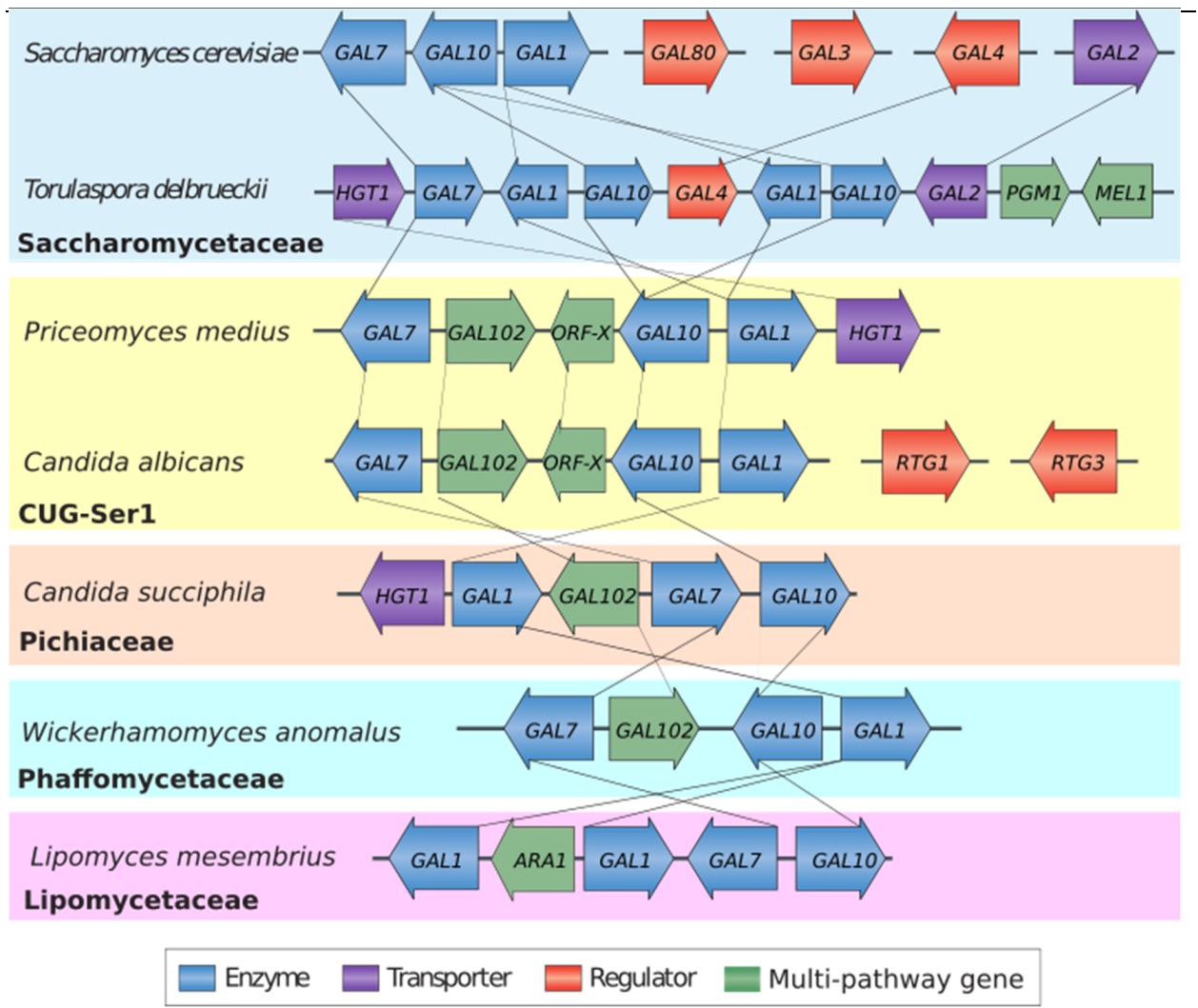

**Figure 2.** Genomic organization of *GAL* gene clusters in different budding yeast major clades. Note the differing patterns of presence / absence of *GAL* pathway genes in between major clades (indicated by the large colored rectangles). Lines correspond to homologs; gene box colors correspond to different functional categories of genes. Information displayed in the figure based on: Shen et al. 2018, Haase et al. 2021, and Venkatesh et al. 2021.

---

The variety of budding yeast clusters shows that functional *GAL* pathways can evolve into several different organizations in the genome. One prominent example is the giant *GAL* clusters in *Torulaspora* species, where *GAL4, MEL1, GAL2*, *PGM1/2*, and *HGT1*, as well as multiple copies of *GAL1* and *GAL10* are parts of the same cluster (Venkatesh et al. 2020, Figure 2). These additional components of the cluster are likely functionally significant in environments with high amounts of melibiose or galactose: Mel1p breaks down melibiose into galactose and glucose, Gal2p transports galactose into the cell, Gal4p can upregulate transcription of the pathway, Hgt1p can transport glucose (and possibly other sugars) generated by the pathway, and Pgm1/2p converts the glucose-1-phosphate generated by the Leloir pathway to glucose-6-phosphate, which can then go through glycolysis. Interestingly, the clustering of these genes with *GAL1, GAL7,* and *GAL10* is not observed in other budding yeasts, with the notable exception of *HGT1*, which is frequently found in *GAL* clusters in the CUG-Ser1 clade (e.g., in *Priceomyces medius* – see Figure 2).

The clustering of the *GAL* genes makes the *GAL* cluster genomic region a good candidate for HGT since acquisition of the *GAL* cluster would provide the minimal genetic information necessary for the utilization of the available galactose in the environment. Acquisition of the *GAL* cluster could occur in organisms with a functional *GAL* pathway as well as in organisms whose ancestors lost the pathway – there have been many instances of *GAL* pathway loss across the subphylum (Riley et al. 2016, Slot & Rokas 2010, Haase et al. 2021, Wolfe et al. 2015, Hittinger et al. 2004) – suggesting that pathway losses are potentially reversible. Support for this hypothesis comes from a recent molecular phylogenetic study that inferred HGT of a CUG-Ser1 type of *GAL* cluster independently occurred in the genera *Brettanomyces*, *Wickerhamomyces*, and *Nadsonia* after ancestral losses of the *GAL* pathway in at least two lineages (Haase et al. 2021). This inference is supported by the observation that the *GAL* clusters of the recipient species share the same cluster organization as the donor species and by formal topology tests that show their phylogenetic placement to be closer to the CUG-Ser1 clade than to their known evolutionary relatives (Haase et al. 2021, Figure 2).

CUG-Ser1 yeasts have repeatedly served as HGT donors of *GAL* clusters to organisms in other budding yeast major clades, but there are no known instances of *GAL* cluster HGTs from Saccharomycetaceae to lineages outside of the family. It has been hypothesized that this difference is due to the *RTG1/RTG3* mode of regulation used by the CUG-Ser1 clade. These transcription factors are more broadly conserved than *GAL4*, which is not known to regulate the *GAL* genes outside of the family Saccharomycetaceae (Dalal et al. 2016, Ata et al. 2018). In fact, even though *C. albicans* has a *GAL4* ortholog, it is much shorter in length and has been shown to regulate an entirely different set of genes (Martchenko et al. 2007). Similarly, in the Pichiaceae clade, a *GAL4* ortholog has been found to regulate Crabtree-Warburg Effect in *Komagataella phaffii* instead of the *GAL* pathway (Ata et al. 2018). *GAL* clusters in the CUG-Ser1 clade have low background levels of gene expression, whereas clusters in the family Saccharomycetaceae are typically actively repressed in the presence of glucose (Haase et al.

2021). Thus, *GAL* clusters acquired from the CUG-Ser1 clade would be more likely to be basally expressed and less likely to require the evolution of a new mode of regulation in the recipient organisms. Interestingly, CUG-Ser1 yeasts can act as donors for HGT even to lineages outside of budding yeasts; for example, *Schizosaccharomyces* fission yeasts, an independently evolved lineage of yeasts in the subphylum Taphrinomycotina, also acquired their *GAL* cluster via HGT from CUG-Ser1 yeasts (Slot & Rokas, 2010).

**Box 2. Testing the Evolutionary Advantage(s) of *GAL* Gene Clustering**

Whether the clustering of *GAL* (and sometimes other functionally related) genes is evolutionary advantageous remains a major, outstanding question and several genetic (e.g., coordinated gene expression, genetic linkage) and phenotypic (e.g., avoidance of toxic intermediates) models have been proposed to explain its origin and maintenance (Lang et al. 2011, McGary et al. 2013, Xu et al. 2019, Rokas et al. 2018). For example, it has been previously observed that metabolic genes in pathways with toxic intermediates are more often clustered together than those in pathways lacking them (McGary et al. 2013). In the context of the *GAL* pathway, galactose-1-phosphate is a toxic intermediate, which leads the occurrence of a disease called galactosemia in humans lacking the functional pathway due to their inability to metabolize this intermediate. Furthermore, the metabolic genes encoding the enzymes involved in the production and conversion of a toxic intermediate were most often divergently oriented (an arrangement typical of co-regulated genes), such as *GAL1* and *GAL10* in *C. albicans* and *S. cerevisiae* (Figure 1), suggesting that clustering may be associated with reducing the buildup of the toxic intermediate (McGary et al. 2013). In support of this hypothesis, a recent paper by Xu et al. found that experimental unlinking of the *GAL1, GAL7,* and *GAL10* genes in *S. cerevisiae* leads to higher fluctuations of their expression levels – and presumably higher buildup of the toxic intermediate galactose-1-phosphate – compared to when the three genes are clustered (Xu et al. 2019).

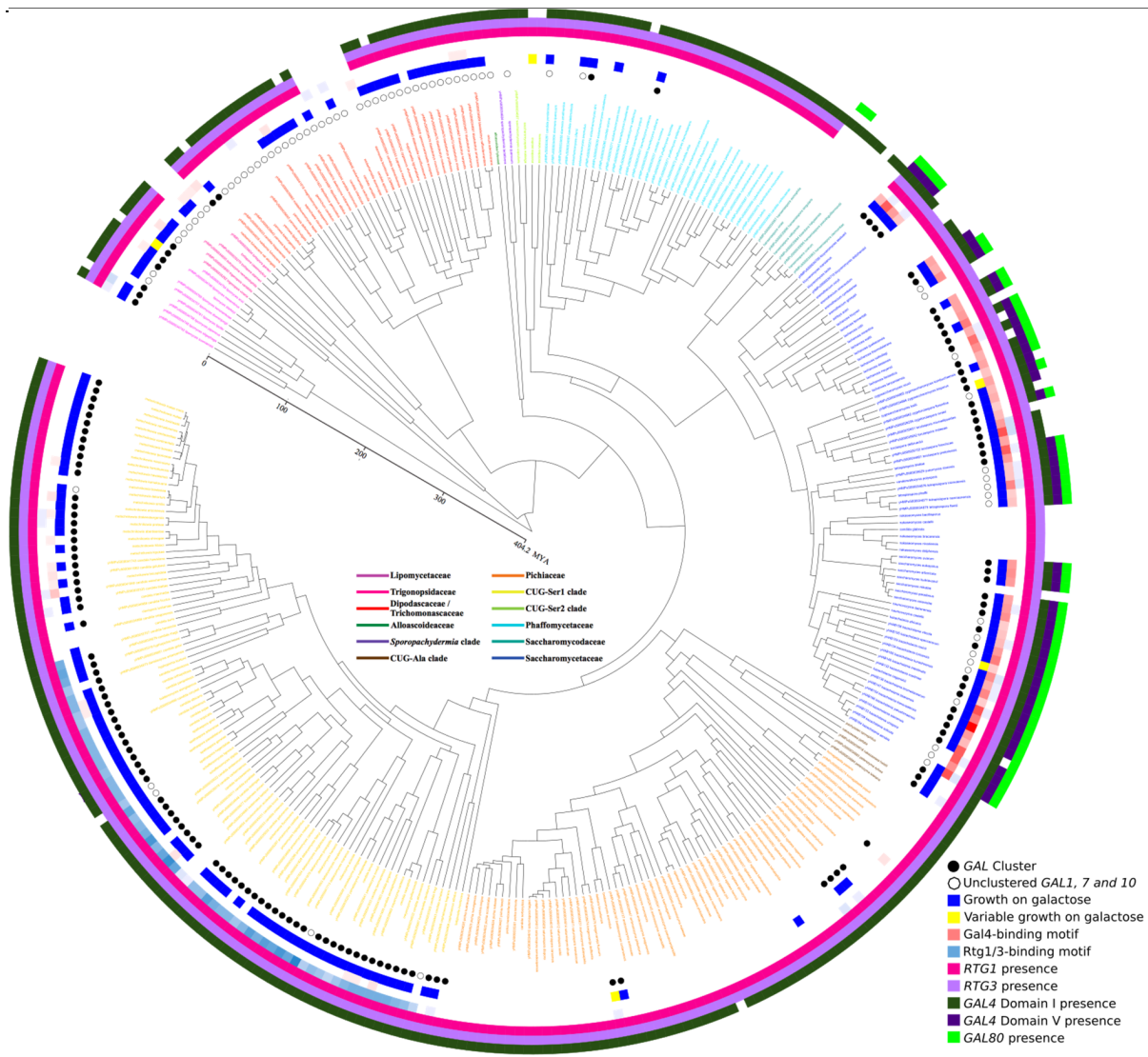

**Figure 3.** Gene presence and absence of key transcription factors involved in *GAL* pathway regulation, as well as variation in the presence and absence of their transcription factor-binding site sequence motifs across budding yeasts. Clustering is defined as *GAL1, GAL7,* and *GAL10* having 0-5 ORFs between them. Domain I of *GAL4* encodes amino acids 1-76 and Domain V amino acids 767-881 of the *S. cerevisiae* protein (Traven et al.

2006, Pan & Coleman 1987). Information displayed in the figure based on Shen et al. 2018, Labella et al. 2021, Opulente et al, 2018.

---

**Evolution of Gene Regulation in the *GAL* Pathway**

While the *GAL4* regulatory system appears to be conserved throughout the family Saccharomycetaceae, there are still significant differences in *GAL* gene induction and repression between species. For example, in 1% glucose medium supplement with galactose, *S. cerevisiae* waits until glucose is completely exhausted to start metabolizing galactose, while the closely related species *Saccharomyces uvarum* does not (Roop et al. 2016). The stronger repression and slower induction of the *GAL* genes in *S. cerevisiae* compared to *S. uvarum* (Roop et al. 2016, Caudy et al. 2013) gives *S. cerevisiae* a fitness advantage in environments where glucose is in excess but a disadvantage when switching from glucose to galactose (Roop et al. 2016). When the promoter or the *GAL* coding regions found in *S. cerevisiae* were expressed in *S. uvarum*, this phenotype could be partially reconstructed, suggesting that both the coding and promoter regions contribute to this *S. cerevisiae* phenotype. Thus, even within the *GAL4* regulatory system, there can be significant differences in induction and repression of the *GAL* genes.

Comparison of the *GAL* pathways of *S. cerevisiae*, *S. uvarum*, and related species has revealed several other differences. Regulatory differences were first noted between *S. uvarum* and *S. cerevisiae* when it was found that the *S. uvarum* genome has retained both *GAL80* and *GAL80B* (the two genes are ohnologs stemming from the ancient whole duplication event, and both were predicted by machine learning to be involved in galactose assimilation), while the *S. cerevisiae* genome has retained only *GAL80* (Caudy et al. 2013, Hittinger et al. 2004). Recent work demonstrated that the corepressor encoded by *GAL80B* in *S. uvarum* plays an important role in preventing metabolic overload when grown in galactose, especially when alternative sugars are also being metabolized (Kuang et al. 2016). Consequently, while a *GAL80* knockout in *S. cerevisiae* grew more quickly in galactose, a double knockout of *GAL80* and *GAL80B* caused a growth arrest of *S. uvarum* when grown in galactose. Finally, Gal4p-binding sites upstream of the metabolic bottleneck gene *PGM1* in *S. uvarum* and several other species of the family Saccharomycetaceae lead to significantly faster growth on galactose compared to species lacking these binding sites (either naturally or via knock out experiments) (Kuang et al. 2018). Interestingly, these more active *GAL* networks (Caudy et al. 2013, Roop et al. 2016, Kuang et al. 2016) tend to be found in the same yeast species that also have dual layers of repression (Kuang et al. 2018). Thus, it seems that species are continually dialing their regulatory systems to the availability of galactose and other sugars in their environments.

While the *GAL4* mode of regulation has been well-characterized in *Saccharomyces* species, less is known about the regulation of the *GAL* pathway in other species of budding yeasts, especially those in other major clades. Recently, it was discovered that the heterodimeric HLH transcription factors Rtg1p and Rtg3p regulate the *GAL* pathway of *C. albicans* (Dalal et al. 2016). This regulatory mode differs from the Gal4p mode because it has a higher basal level of expression and a more graded induction, rather than the bimodal, "on-and-off switch" mode of Gal4p regulation. Due to the conservation of *RTG1* across the budding yeast subphylum, as well as reduced induction of *GAL1* in the presence of galactose by a knockout of *RTG1* in the outgroup species *Yarrowia lipolytica,* it was suggested that Rtg1p-Rtg3p mode of regulation may be the ancestral one (Dalal et al. 2016). In support of this hypothesis, *RTG1* and *RTG3* are conserved through most of the budding yeast phylogeny (Figure 3). Domain I encoded by *GAL4* (amino acid residues 1-76, encoding a DNA-binding domain homologous to dozens of transcription factors) is similarly conserved, but Domain V encoded by *GAL4* (residues 767-881, encoding the Gal80p-binding domain) is only conserved in Saccharomycetaceae (excluding some *Torulaspora* species), and its pattern of conservation mirrors that of *GAL80* (Figure 3) (Traven et al. 2006, Pan & Coleman 1989). This suggests that the Gal4p-Gal80p mode of regulation is restricted to the Saccharomycetaceae clade, while the Rtg1p-Rtg3p mode may be more broadly conserved. However, although Rtg1/3p-binding motifs are conserved in most galactose-growing species in the CUG-Ser1 clade, a lineage within the genus *Metschnikowia* and a few other species appear to lack these motifs (Figure 3). Furthermore, neither the Gal4p, nor the Rtg1/3p motifs, are highly enriched in the *GAL* cluster regions of other budding yeast clades (Figure 3). These data raise the hypothesis that there is a variety of modes of regulation of galactose metabolism in the budding yeast subphylum.

**Population-level variation of *GAL* gene clusters**

In recent years, examination of isolates from diverse environments has substantially enhanced our understanding of genomic and phenotypic variation in populations of budding yeast species (Libkind et al. 2020, Hénault et al. 2019, Peter et al. 2018). While the genomic organization and regulation of *GAL* gene clusters vary between budding yeast species, numerous population genomic studies have shown that *GAL* function and regulation can also vary substantially within species. The first example of substantial population-level variation in the *GAL* gene cluster was described in *Saccharomyces kudriavzevii*, a close relative of *S. cerevisiae*. Whereas European isolates of *S. kudriavzevii* have a functional *GAL* pathway comprised of six genes (compared to *S.*

*cerevisiae*, they only lack the optional co-inducer encoded by *GAL3*) and can grow on galactose, Eastern Asian isolates cannot grow on galactose and their *GAL* pathway is composed of pseudogenes that are syntenic with the functional alleles (Hittinger et al. 2010). If a crossing occurred between these populations, meiotic progeny would be unlikely to harbor either completely functional or completely non-functional networks, and some of the partial networks are less fit than either parent. For example, *S. kudriavzevii* segregants that lacked the corepressor encoded by *GAL80* but had other functional genes would constitutively express them in environments that lack galactose, often at a substantial fitness cost (Hittinger et al. 2010). This example shows that the *GAL* genes of budding yeast populations can adapt to different environments and regulatory systems, including maintaining completely non-functional versions as local adaptations and balanced polymorphisms.

Recent examinations of the genomes of more than one thousand *S. cerevisiae* isolates have revealed the existence of three different combinations of highly divergent alleles in their *GAL* gene networks (Boocock et al. 2021, Legras et al. 2018, Duan et al. 2019, Duan et al, 2018), suggesting that substantial population-level variation in the *GAL* pathway may be more common than previously thought in yeast populations. The first combination, found in most isolates, is composed of the alleles found in the reference S288C strain of *S. cerevisiae.* The second combination, found in a small percentage of isolates from different environments, including dairy ones, is composed of highly diverged alleles in the *GAL2, GAL1/7/10*, and *PGM1* genes; these alleles' divergence from their reference counterparts predates both the origin of the species *S. cerevisiae*, as well as the origin of the genus *Saccharomyces* (Legras et al. 2018, Boocock et al. 2021, Duan et al. 2019). The third combination, found in a few Chinese isolates from soil or wood environments, is composed of the same highly diverged *GAL1/7/10* alleles as the second combination, but the alleles for the *GAL2* and *PGM1* loci differ substantially from those found in the other two combinations; these alleles' origin(s) also predates the origin of the genus *Saccharomyces* (Boocock et al. 2021, Duan et al. 2019).

The common characteristic of these alternative combinations of *GAL* gene network variants is that they allow these isolates to grow faster on galactose and slower on glucose than the reference S288C strain (which contains the first combination). For example, the *PGM1* allele of some of these isolates has a Gal4p-regulated promoter, allowing for quick utilization of galactose for energy through glycolysis, while the *PGM1* allele of the reference strain does not; this allele is incompatible with the *GAL2* and *GAL1/7/10* alleles of isolates with the second combination (Boocock et al. 2021). Similarly, mutations in certain isolates have impaired or abolished Mig1p- and Gal80p-mediated repression of the *GAL* pathway (Duan et al. 2019). Finally, these isolates show – in addition to the observed variation in their *GAL2* loci – extensive variation in their hexose transporters, with numerous examples of gene fusion events, gene truncations, and wholesale gene deletions (Duan et al. 2019, Legras et al. 2018).

The very deep origin of the observed variation in the *GAL* gene network of *S. cerevisiae* raises the question of the evolutionary processes involved in its making and maintenance. Two alternative hypotheses have been proposed: introgression or HGT from a yet-to-be-identified species closely related to the genus *Saccharomyces* (Legras et al. 2018, Duan et al. 2019) and ancient balancing selection (Boocock et al. 2021); a combination of these two evolutionary scenarios (e.g., introgression followed by balancing selection) is also a possibility. Extensive genetic simulations reject a scenario where the alternative variants stem from a recent introgression, perhaps around the beginnings of agriculture, favoring instead a scenario of ancient balancing selection or ancient introgression followed by balancing selection (Boocock et al. 2021).

Recent studies have begun to identify similar types of adaptations in other budding yeast metabolic genes and pathways. For example, it was recently shown that *Kluyveromyces lactis* var. *lactis*, the dairy-based variety of *K. lactis,* acquired the ability to ferment the milk sugar lactose via a recent HGT of a *LAC4-LAC12* gene cluster from a dairy isolate of *Kluyveromyces marxianus* (Valera et al. 2019). In this cluster, *LAC4* encodes the enzyme lactase that hydrolyzes lactose into galactose and glucose, and *LAC12* encodes the transporter lactose permease (Valera et al. 2019).

These population genomic examinations raise the hypothesis that the *GAL* pathway, and perhaps most other metabolic pathways, of budding yeasts are subject to varied selection for the utilization of the galactose present in different environments and conditions. Thus, their pathways are likely to harbor multiple, highly divergent gene network variants and exhibit strong signatures of local adaptation. If this turns out to be true, population genomic examinations of populations of the ~1,200 known species of budding yeasts could reveal a treasure-trove of novel *GAL* (and other metabolic) gene network variants that would not only enhance our understanding of eukaryotic pathway evolution, but also of yeast metabolic engineering.

**Concluding Remarks**

The *GAL* pathway of the budding yeast *S. cerevisiae* is a favorite textbook example of the regulation of gene expression in eukaryotes (e.g., Griffths et al. 2020). Recent advances in genome sequencing of budding yeasts, coupled with evolutionary genomic and functional studies of diverse populations and species across the subphylum, have allowed for many significant insights on the evolution of the *GAL* pathway that extend far beyond the insights obtained from the classic work in *S. cerevisiae* (and more recently in *C. albicans*). Since this pathway has long served as a model for gene regulation and evolution in eukaryotes, these insights are crucial for broadening our understanding of how metabolic pathways, and their mode of regulation, change in response to different environments. Nevertheless, several questions remain (see Outstanding Questions). The addition of more genomes from diverse budding yeast species and populations found in different environments, coupled with the development of new genetic model systems, is likely to reveal new and interesting findings. Fleshing out the *GAL* pathway's evolutionary and ecological diversity will help bridge our understanding of how genotypic variation emerges as phenotypic variation, perpetuating the pathway's utility as a model.

---

**Box 3. Outstanding Questions**

- How often does HGT of the *GAL* pathway occur in budding yeasts?
- What is the contribution of the observed extensive segregating variation on *GAL* genes in budding yeast populations to the macroevolution of the *GAL* pathway?
- What, if any, is the function of *GAL102* in the pathway?
- How many times has similar clustering of galactose metabolism genes independently evolved in budding yeasts, and what evolutionary pressures could be driving this adaptation (what do these yeast species have in common)?

- How do clustering and regulation affect HGT and maintenance of balanced polymorphisms?
- What regulates galactose metabolism in budding yeast lineages that lack known transcription factors and/or their biding sites but have *GAL* genes and grow on galactose (e.g., in yeasts of the genus *Metschnikowia* or *Lipomyces*)?
- Can we infer with any degree of confidence the genomic organization and regulation of the *GAL* pathway of the budding yeast common ancestor or at various key nodes?
- Is the *GAL* pathway in budding yeasts involved in glycosylation?

---

**Acknowledgements**

We thank members of the Rokas and Hittinger labs, especially members of the Y1000+ Project team, for the helpful suggestions. M-C. H. was supported by the Biological Sciences Graduate Program at Vanderbilt University. Research in C.T.H's lab is supported by National Science Foundation under Grant No. DEB-1442148, in part by the DOE Great Lakes Bioenergy Research Center (DOE BER Office of Science DE-SC0018409), the USDA National Institute of Food and Agriculture (Hatch Project 1020204), a Pew Scholarship in the Biomedical Sciences (Pew Charitable Trusts), and an H. I. Romnes Faculty Fellowship (Office of the Vice Chancellor for Research and Graduate Education with funding from the Wisconsin Alumni Research Foundation). Research in A.R.'s lab is supported by the National Science Foundation (DEB-1442113), the Burroughs Wellcome Fund, and the National Institutes of Health/National Institute of Allergy and Infectious Diseases (R56AI146096).

**References:**

1. Acosta, P. B., & Gross, K. C. (1995). Hidden sources of galactose in the environment. *European Journal of Pediatrics*, *154*(7 Suppl 2), S87-92. https://doi.org/10.1007/BF02143811

2. Coelho, A. I., Berry, G. T., & Rubio-Gozalbo, M. E. (2015). Galactose metabolism and health. *Current Opinion in Clinical Nutrition & Metabolic Care*, *18*(4), 422–427. https://doi.org/10.1097/MCO.0000000000000189

3. Sellick, C. A., Campbell, R. N., & Reece, R. J. (2008). Galactose metabolism in yeast-structure and regulation of the Leloir pathway enzymes and the genes encoding them. *International Review of Cell and Molecular Biology*, *269*, 111–150. https://doi.org/10.1016/S1937-6448(08)01003-4

4. Johnston, M. (1987). A model fungal gene regulatory mechanism: The *GAL* genes of *Saccharomyces cerevisiae*. *Microbiology and Molecular Biology Reviews*, *51*(4), 458–476.

5. Hittinger, C. T., Rokas, A., & Carroll, S. B. (2004). Parallel inactivation of multiple *GAL* pathway genes and ecological diversification in yeasts. Proceedings of the National Academy of Sciences, 101(39), 14144–14149. https://doi.org/10.1073/pnas.0404319101

6. Hittinger, C. T., Gonçalves, P., Sampaio, J. P., Dover, J., Johnston, M., & Rokas, A. (2010). Remarkably ancient balanced polymorphisms in a multi-locus gene network. *Nature*, *464*(7285), 54–58. https://doi.org/10.1038/nature08791

7. Slot, J. C., & Rokas, A. (2010). Multiple *GAL* pathway gene clusters evolved independently and by different mechanisms in fungi. *Proceedings of the National Academy of Sciences*, *107*(22), 10136–10141. https://doi.org/10.1073/pnas.0914418107

8. Dalal, C. K., Zuleta, I. A., Mitchell, K. F., Andes, D. R., El-Samad, H., & Johnson, A. D. (2016). Transcriptional rewiring over evolutionary timescales changes quantitative and qualitative properties of gene expression. *ELife*, *5*. https://doi.org/10.7554/eLife.18981

9. Reily, C., Stewart, T. J., Renfrow, M. B., & Novak, J. (2019). Glycosylation in health and disease. *Nature Reviews Nephrology*, *15*(6), 346–366. https://doi.org/10.1038/s41581-019-0129-4

10. Sasaoka, N., Imamura, H., & Kakizuka, A. (2018). A Trace Amount of Galactose, a Major Component of Milk Sugar, Allows Maturation of Glycoproteins during Sugar Starvation. *IScience*, *10*, 211–221. https://doi.org/10.1016/j.isci.2018.11.035

11. Van Ende, M., Wijnants, S., & Van Dijck, P. (2019). Sugar Sensing and Signaling in *Candida albicans* and *Candida glabrata*. *Frontiers in Microbiology*, *10*. https://doi.org/10.3389/fmicb.2019.00099

12. Brown, V., Sabina, J., & Johnston, M. (2009). Specialized Sugar Sensing in Diverse Fungi. *Current Biology*, *19*(5), 436–441. https://doi.org/10.1016/j.cub.2009.01.056

13. Shen, X.-X., Opulente, D. A., Kominek, J., Zhou, X., Steenwyk, J. L., Buh, K. V., Haase, M. A. B., Wisecaver, J. H., Wang, M., Doering, D. T., Boudouris, J. T., Schneider, R. M., Langdon, Q. K., Ohkuma, M., Endoh, R., Takashima, M., Manabe, R., Čadež, N., Libkind, D., … Rokas, A. (2018). Tempo and Mode of Genome Evolution in the Budding Yeast Subphylum. *Cell*, *175*(6), 1533-1545.e20. https://doi.org/10.1016/j.cell.2018.10.023

14. Shen, X.-X., Steenwyk, J. L., LaBella, A. L., Opulente, D. A., Zhou, X., Kominek, J., Li, Y., Groenewald, M., Hittinger, C. T., & Rokas, A. (2020). Genome-scale phylogeny and contrasting modes of genome evolution in the fungal phylum Ascomycota. *Science Advances*, *6*(45). https://doi.org/10.1126/sciadv.abd0079

15. Li, Y., Steenwyk, J. L., Chang, Y., Wang, Y., James, T. Y., Stajich, J. E., Spatafora, J. W., Groenewald, M., Dunn, C. W., Hittinger, C. T., Shen, X.-X., & Rokas, A. (2021). A genome-scale phylogeny of the kingdom Fungi. *Current Biology: CB*, *31*(8), 1653-1665.e5. https://doi.org/10.1016/j.cub.2021.01.074

16. Kurtzman, C., Fell, J. W., & Boekhout, T. (2011). *The Yeasts: A Taxonomic Study*. Elsevier.

17. Riley, R., Haridas, S., Wolfe, K. H., Lopes, M. R., Hittinger, C. T., Göker, M., Salamov, A. A., Wisecaver, J. H., Long, T. M., Calvey, C. H., Aerts, A. L., Barry, K. W., Choi, C., Clum, A., Coughlan, A. Y., Deshpande, S., Douglass, A. P., Hanson, S. J., Klenk, H.-P., … Jeffries, T. W. (2016). Comparative genomics of biotechnologically important yeasts. *Proceedings of the National Academy of Sciences*, *113*(35), 9882–9887. https://doi.org/10.1073/pnas.1603941113

18. Krassowski, T., Coughlan, A. Y., Shen, X.-X., Zhou, X., Kominek, J., Opulente, D. A., Riley, R., Grigoriev, I. V., Maheshwari, N., Shields, D. C., Kurtzman, C. P., Hittinger, C. T., Rokas, A., & Wolfe, K. H. (2018). Evolutionary instability of CUG-Leu in the genetic code of budding yeasts. *Nature Communications*, *9*(1), 1887. https://doi.org/10.1038/s41467-018-04374-7

19. Opulente, D. A., Rollinson, E. J., Bernick-Roehr, C., Hulfachor, A. B., Rokas, A., Kurtzman, C. P., & Hittinger, C. T. (2018). Factors driving metabolic diversity in the budding yeast subphylum. *BMC Biology*, *16*(1), 26. https://doi.org/10.1186/s12915-018-0498-3

20. LaBella, A. L., Opulente, D. A., Steenwyk, J. L., Hittinger, C. T., & Rokas, A. (2021). Signatures of optimal codon usage in metabolic genes inform budding yeast ecology. *PLOS Biology*, *19*(4), e3001185. https://doi.org/10.1371/journal.pbio.3001185

21. Wolfe, K. H., & Shields, D. C. (1997). Molecular evidence for an ancient duplication of the entire yeast genome. *Nature*, *387*(6634), 708–713. https://doi.org/10.1038/42711

22. Marcet-Houben, M., & Gabaldón, T. (2015). Beyond the Whole-Genome Duplication: Phylogenetic Evidence for an Ancient Interspecies Hybridization in the Baker's Yeast Lineage. *PLoS Biology*, *13*(8), e1002220. https://doi.org/10.1371/journal.pbio.1002220

23. Haase, M. A. B., Kominek, J., Opulente, D. A., Shen, X.-X., LaBella, A. L., Zhou, X., DeVirgilio, J., Hulfachor, A. B., Kurtzman, C. P., Rokas, A., & Hittinger, C. T. (2021). Repeated horizontal gene transfer of *GAL*actose metabolism genes violates Dollo's law of irreversible loss. *Genetics*, *217*(iyaa012). https://doi.org/10.1093/genetics/iyaa012

24. Benocci, T., Aguilar-Pontes, M. V., Kun, R. S., Seiboth, B., Vries, R. P. de, & Daly, P. (2018). *ARA1* regulates not only l-arabinose but also D-galactose catabolism in *Trichoderma reesei*. *FEBS Letters*, *592*(1), 60–70. https://doi.org/10.1002/1873-3468.12932

25. Venkatesh, A., Murray, A. L., Coughlan, A. Y., & Wolfe, K. H. (2021). Giant *GAL* gene clusters for the melibiose-galactose pathway in *Torulaspora*. *Yeast*, *38*(1), 117–126. https://doi.org/10.1002/yea.3532

26. Martchenko, M., Levitin, A., & Whiteway, M. (2007). Transcriptional activation domains of the *Candida albicans* Gcn4p and Gal4p homologs. *Eukaryotic Cell*, *6*(2), 291–301. https://doi.org/10.1128/EC.00183-06

27. Ata, Ö., Rebnegger, C., Tatto, N. E., Valli, M., Mairinger, T., Hann, S., Steiger, M. G., Çalık, P., & Mattanovich, D. (2018). A single Gal4-like transcription factor activates the Crabtree effect in *Komagataella phaffii*. *Nature Communications*, *9*(1), 4911. https://doi.org/10.1038/s41467-018-07430-4

28. Lang, G. I., & Botstein, D. (2011). A test of the coordinated expression hypothesis for the origin and maintenance of the *GAL* cluster in yeast. *PloS One*, *6*(9), e25290. https://doi.org/10.1371/journal.pone.0025290

29. McGary, K. L., Slot, J. C., & Rokas, A. (2013). Physical linkage of metabolic genes in fungi is an adaptation against the accumulation of toxic intermediate compounds. *Proceedings of the National Academy of Sciences*, *110*(28), 11481–11486. https://doi.org/10.1073/pnas.1304461110

30. Xu, H., Liu, J.-J., Liu, Z., Li, Y., Jin, Y.-S., & Zhang, J. (2019). Synchronization of stochastic expressions drives the clustering of functionally related genes. *Science Advances*, *5*(10), eaax6525. https://doi.org/10.1126/sciadv.aax6525

31. Rokas, A., Wisecaver, J. H., & Lind, A. L. (2018). The birth, evolution and death of metabolic gene clusters in fungi. *Nature Reviews Microbiology*, *16*(12), 731–744. https://doi.org/10.1038/s41579-018-0075-3

32. Caudy, A. A., Guan, Y., Jia, Y., Hansen, C., DeSevo, C., Hayes, A. P., Agee, J., Alvarez-Dominguez, J. R., Arellano, H., Barrett, D., Bauerle, C., Bisaria, N., Bradley, P. H., Breunig, J. S., Bush, E., Cappel, D., Capra, E., Chen, W., Clore, J., … Dunham, M. J. (2013). A New System for Comparative Functional Genomics of *Saccharomyces* Yeasts. *Genetics*, *195*(1), 275–287. https://doi.org/10.1534/genetics.113.152918

33. Kuang, M. C., Hutchins, P. D., Russell, J. D., Coon, J. J., & Hittinger, C. T. (2016). Ongoing resolution of duplicate gene functions shapes the diversification of a metabolic network. *ELife*, *5*. https://doi.org/10.7554/eLife.19027

34. Kuang, M. C., Kominek, J., Alexander, W. G., Cheng, J.-F., Wrobel, R. L., & Hittinger, C. T. (2018). Repeated Cis-Regulatory Tuning of a Metabolic Bottleneck Gene during Evolution. *Molecular Biology and Evolution*, *35*(8), 1968–1981. https://doi.org/10.1093/molbev/msy102

35. Roop, J. I., Chang, K. C., & Brem, R. B. (2016). Polygenic evolution of a sugar specialization trade-off in yeast. *Nature*, *530*(7590), 336–339. https://doi.org/10.1038/nature16938

36. Pan, T., & Coleman, J. E. (1989). Structure and function of the Zn(II) binding site within the DNA-binding domain of the GAL4 transcription factor. *Proceedings of the National Academy of Sciences of the United States of America*, *86*(9), 3145–3149.

37. Pan, T., & Coleman, J. E. (1989). Structure and function of the Zn(II) binding site within the DNA-binding domain of the GAL4 transcription factor. *Proceedings of the National Academy of Sciences of the United States of America*, *86*(9), 3145–3149.

38. Libkind, D., Peris, D., Cubillos, F. A., Steenwyk, J. L., Opulente, D. A., Langdon, Q. K., Rokas, A., & Hittinger, C. T. (2020). Into the wild: New yeast genomes from natural environments and new tools for their analysis. *FEMS Yeast Research*, *20*(2). https://doi.org/10.1093/femsyr/foaa008

39. Hénault, M., Eberlein, C., Charron, G., Durand, É., Nielly-Thibault, L., Martin, H., & Landry, C. R. (2019). Yeast Population Genomics Goes Wild: The Case of Saccharomyces paradoxus. In M. F. Polz & O. P. Rajora (Eds.), *Population Genomics: Microorganisms* (pp. 207–230). Springer International Publishing. https://doi.org/10.1007/13836_2017_4

40. Peter, J., De Chiara, M., Friedrich, A., Yue, J.-X., Pflieger, D., Bergström, A., Sigwalt, A., Barre, B., Freel, K., Llored, A., Cruaud, C., Labadie, K., Aury, J.-M., Istace, B., Lebrigand, K., Barbry, P., Engelen, S., Lemainque, A., Wincker, P., … Schacherer, J. (2018). Genome evolution across 1,011 *Saccharomyces cerevisiae* isolates. *Nature*, *556*(7701), 339–344. https://doi.org/10.1038/s41586-018-0030-5

41. Boocock, J., Sadhu, M. J., Durvasula, A., Bloom, J. S., & Kruglyak, L. (2021). Ancient balancing selection maintains incompatible versions of the galactose pathway in yeast. *Science*, *371*(6527), 415–419. https://doi.org/10.1126/science.aba0542

42. Duan, S.-F., Shi, J.-Y., Yin, Q., Zhang, R.-P., Han, P.-J., Wang, Q.-M., & Bai, F.-Y. (2019). Reverse Evolution of a Classic Gene Network in Yeast Offers a Competitive Advantage. *Current Biology: CB*, *29*(7), 1126-1136.e5. https://doi.org/10.1016/j.cub.2019.02.038

43. Duan, S.-F., Han, P.-J., Wang, Q.-M., Liu, W.-Q., Shi, J.-Y., Li, K., Zhang, X.-L., & Bai, F.-Y. (2018). The origin and adaptive evolution of domesticated populations of yeast from Far East Asia. *Nature Communications*, *9*(1), 2690. https://doi.org/10.1038/s41467-018-05106-7

44. Legras, J.-L., Galeote, V., Bigey, F., Camarasa, C., Marsit, S., Nidelet, T., Sanchez, I., Couloux, A., Guy, J., Franco-Duarte, R., Marcet-Houben, M., Gabaldon, T., Schuller, D., Sampaio, J. P., & Dequin, S. (2018). Adaptation of *S. cerevisiae* to Fermented Food Environments Reveals Remarkable Genome Plasticity and the Footprints of Domestication. *Molecular Biology and Evolution*, *35*(7), 1712–1727. https://doi.org/10.1093/molbev/msy066

45. Varela, J. A., Puricelli, M., Ortiz-Merino, R. A., Giacomobono, R., Braun-Galleani, S., Wolfe, K. H., & Morrissey, J. P. (2019). Origin of Lactose Fermentation in *Kluyveromyces lactis* by Interspecies Transfer of a Neo-functionalized Gene Cluster during Domestication. *Current Biology*, *29*(24), 4284-4290.e2. https://doi.org/10.1016/j.cub.2019.10.044

46. Griffiths, A. J. F., Doebley, J., Peichel, C., & Wassarman, D. A. (2020). *An Introduction to Genetic Analysis*. Macmillan Learning.